\documentclass[journal,twoside,web]{ieeecolor}
\usepackage{generic}
\usepackage{cite}
\usepackage{amsmath,amssymb,amsfonts}
\usepackage{algorithmic}
\usepackage{graphicx}
\usepackage{textcomp}
\usepackage{hyperref}
\usepackage{rotating}
\usepackage[outdir=./]{epstopdf}
\usepackage{array,bigstrut,makecell,multirow,rotating}
\def\BibTeX{{\rm B\kern-.05em{\sc i\kern-.025em b}\kern-.08em
    T\kern-.1667em\lower.7ex\hbox{E}\kern-.125emX}}
\begin{document}
\title{Segmentation of Anatomical Layers and Imaging Artifacts in Intravascular Polarization Sensitive Optical Coherence Tomography Using Attending Physician and Boundary Cardinality Losses}
\author{Mohammad Haft-Javaherian, Martin Villiger, Kenichiro Otsuka, Joost Daemen, Peter Libby, \\ Polina Golland*, and Brett E. Bouma*
\thanks{This work was supported by the National Institutes of Health NIBIB under Grants P41EB-015902 and P41EB-015903, the National Heart, Lung, and Blood Institute Grant 1R01HL134892, the American Heart Association Grant 18CSA34080399), the RRM Charitable Fund, the Simard Fund, and Novartis. Also, Mohammad Haft-Javaherian was supported by Bullock Postdoctoral Fellowship. This work was done partially using MIT-IBM Satori hardware resource. (Corresponding author: Mohammad Haft-Javaherian).}
\thanks{* Polina Golland and Brett E. Bouma have equal contributions.}
\thanks{Mohammad Haft-Javaherian is with the Wellman Center for Photomedicine, Massachusetts General Hospital, Harvard Medical School, Boston, MA 02114 USA, and the Computer Science and Artificial Intelligence Laboratory (CSAIL), Massachusetts Institute of Technology, Cambridge, MA 02142 USA (e-mail: haft@csail.mit.edu).}
\thanks{Martin Villiger and Kenichiro Otsuka are with the Wellman Center for Photomedicine, Massachusetts General Hospital, Harvard Medical School, Boston, MA 02114 USA (e-mail: mvilliger@mgh.harvard.edu; otsukakenichiro1@gmail.com).}
\thanks{Joost Daemenis is with the Department of Interventional Cardiology,
Thoraxcenter, Erasmus Medical Center, Rotterdam, The Netherlands
(e-mail: j.daemen@erasmusmc.nl).}
\thanks{Peter Libby is with the Division of Cardiovascular Medicine, Department of Medicine, Brigham and Women’s Hospital, Harvard Medical School, Boston, MA 02115 (e-mail: plibby@bwh.harvard.edu).}
\thanks{Polina Golland is with the Computer Science and Artificial Intelligence Laboratory (CSAIL), Massachusetts Institute of Technology, Cambridge, MA 02142 USA (e-mail: polina@csail.mit.edu).}
\thanks{Brett E. Bouma is with the Wellman Center for Photomedicine, Massachusetts General Hospital, Harvard Medical School, Boston, MA 02114 USA, and the Institute for Medical Engineering and Science, Massachusetts Institute of Technology, Cambridge, MA 02142 USA (e-mail: bouma@mgh.harvard.edu).}
}

\maketitle

\begin{abstract} 
Intravascular ultrasound and optical coherence tomography are widely available for assessing coronary stenoses and provide critical information to optimize percutaneous coronary intervention. 
Intravascular polarization-sensitive optical coherence tomography (PS-OCT) measures the polarization state of the light scattered by the vessel wall in addition to conventional cross-sectional images of subsurface microstructure.
This affords reconstruction of tissue polarization properties and reveals improved contrast between the layers of the vessel wall along with insight into collagen and smooth muscle content. 
Here, we propose a convolutional neural network model, optimized using a new multi-term loss function, 
that takes advantage of the additional polarization contrast and classifies the lumen, intima, and media layers in addition to guidewire and plaque shadows. 
Our model segments the media boundaries through fibrotic plaques and continues to estimate the outer media boundary behind shadows of lipid-rich plaques.
We demonstrate that our multi-class classification model outperforms existing methods that exclusively use conventional OCT data, predominantly segment the lumen, and consider subsurface layers at most in regions of minimal disease. 
Segmentation of all anatomical layers throughout diseased vessels may facilitate stent sizing and will enable automated characterization of plaque polarization properties for investigation of the natural history and significance of coronary atheromas. 
The source code and the trained model are publicly available at \url{https://github.com/mhaft/OCTseg}.
\end{abstract} 


\begin{figure}
\centerline{\includegraphics[width=0.48\textwidth]{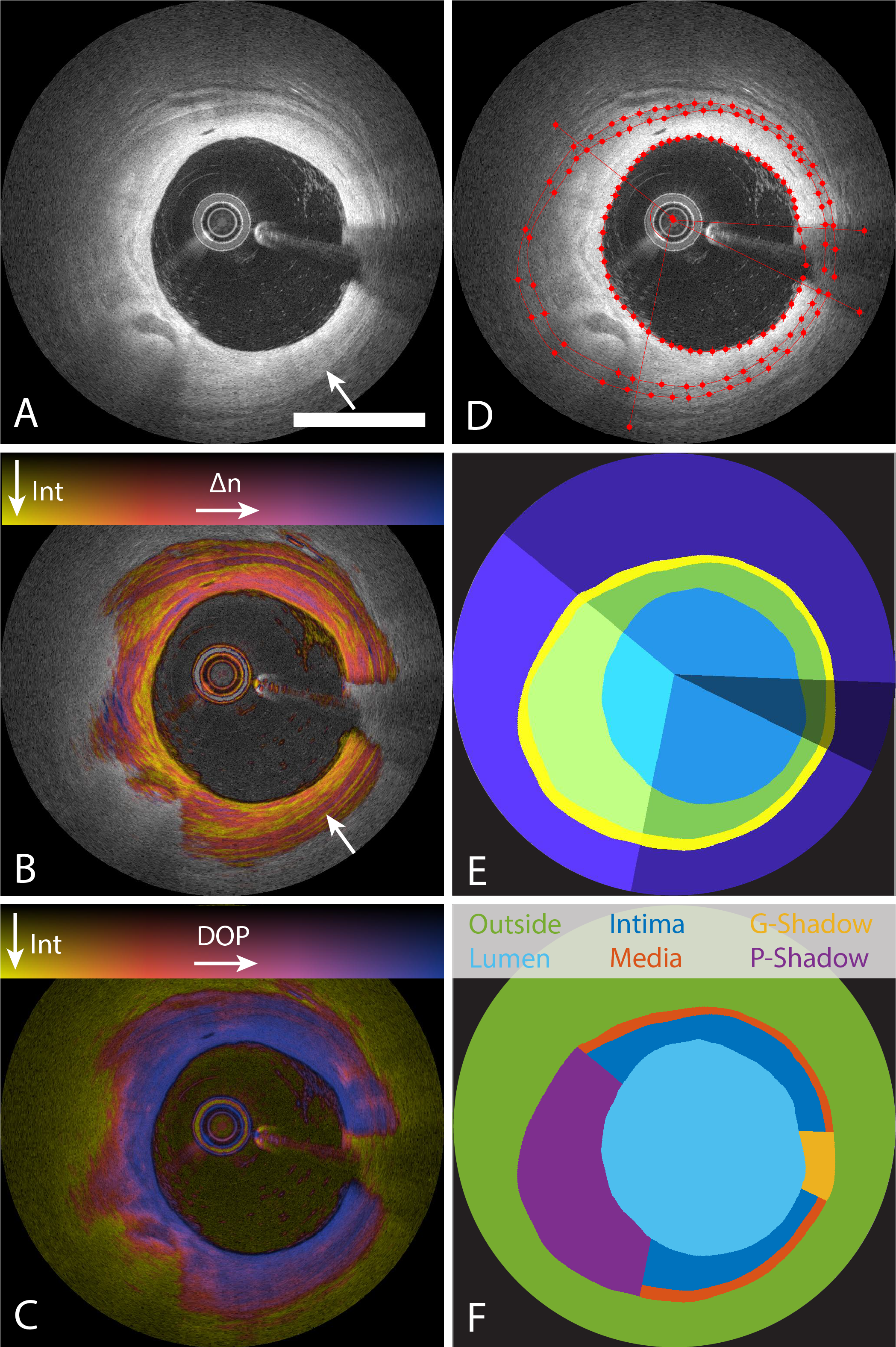}}
\caption{Expert annotation of IV PS-OCT images. 
An example of a multi-channel IV PS-OCT cross-sectional image, including backscatter signal intensity (A), birefringence (B), and depolarization (C) channels. The white arrow highlights a section of the media that exhibits little contrast in the intensity signal, but is readily identified in the birefringence image.
D. Manual annotation by an expert using our Matlab graphical user interface. 
E. Inclusive pixel-level labels derived from the manual annotation (see Data Section). 
F. Equivalent exclusive labels defined in Table~\ref{tab:class_label_def}. 
Scale bar: 2 mm.}
\label{fig:annotations}
\end{figure}

\section{Introduction} \label{sec:introduction} 
\IEEEPARstart{D}{espite} progress with effective therapies for treating acute coronary events, their prediction and prevention continues to present a major clinical challenge~\cite{franco2011challenges}. 
More than one million individuals suffered from acute coronary events last year in the United States alone~\cite{virani2020heart}. 
In addition to pharmacological medical therapy, patients suffering an acute coronary event frequently receive percutaneous coronary intervention (PCI). 
PCI is similarly used in patients with chronic coronary syndrome. 

Intravascular (IV) optical coherence tomography (OCT) is increasingly used for guiding PCI. 
It acquires high resolution images of the subsurface microstructure of coronary atherosclerotic lesions~\cite{jang2001visualization,okamura2011first} and helps with identification of the culprit lesion, stent sizing, and confirming stent implantation. 
The use of IV-OCT achieves better physiological outcomes than using coronary angiography alone~\cite{meneveau2016optical, ali2016optical }, and can assess functional stenosis severity more accurately than intravascular ultrasound (IVUS)~\cite{ramasamy2020optical}. 
Nonetheless, clinical adoption of IV-OCT has been modest~\cite{ali2021intracoronary}. 
This contrasts with its important role as an essential clinical research instrument for investigating the pathophysiology genesis of coronary atherosclerosis~\cite{bouma2017intravascular}. 
One factor impairing more widespread use may be the image contrast, signal statistics, and speckle characteristics specific to OCT, which complicate image interpretation. 
Extensive training affords refined interpretation of pullback data sets but is based nearly exclusively on subjective criteria that are difficult to learn, which results in modest intra- and inter-reader agreement ~\cite{manfrini2006sources}. 
Furthermore, the clinical workflow precludes time-consuming interpretation, emphasizing the need for automated analysis that presents the operator with clear indications to guide the intervention. 
However, until the recent introduction of Ultreon by Abbott~\cite{januszek2022procedure}, image processing algorithms used in the catheterization laboratories have primarily been limited to lumen segmentation, disregarding subsurface vessel morphology. 
The automatic detection of anatomical layers within the vessel wall and other features beyond the lumen promises to refine guidance of PCI and simplify translational research in the catheterization laboratory by eliminating the need for extensive training and time-consuming manual segmentation.

In parallel to conventional OCT (Fig.~\ref{fig:annotations}.A), polarization sensitive (PS) OCT performs polarimetric measurements to reconstruct images of tissue birefringence and depolarization (Fig.~\ref{fig:annotations}.B-C). 
Microscopic PS-OCT in human aortic plaques reports tissue birefringence that can quantify collagen and smooth muscle cell content features, which play an important role in plaque stability and vascular healing~\cite{nadkarni2007measurement}. 
Moreover, catheter-based PS-OCT perform intravascular polarimetry of coronary atherosclerosis~\cite{villiger2018coronary}. 
In addition to tissue birefringence, intravascular polarimetry also measures depolarization, which is increased in tissues containing lipid particles, macrophage accumulations, or cholesterol crystals, as confirmed by correlation with histology in a human cadaver heart study~\cite{villiger2018coronary}. 
Intravascular polarimetry is compatible with current intravascular imaging catheters, which facilitated its translation into the clinic~\cite{villiger2018repeatability, otsuka2020intravascular}. 

Imaging in patients confirmed the improved image contrast available to polarimetric measurements. 
In particular, the smooth muscle cell-rich tunica media features consistent and high birefringence, often separated from the adjacent intima and adventitia layers by fine bands of low birefringence, co-locating with the internal and elastic laminae (IEL, EEL). 
Depolarization enhances lipid-rich lesions and simplifies their differentiation from calcification, which can have similar appearance in conventional intensity tomograms. 
The ability to consistently measure the EEL diameter along the entire coronary would be highly relevant for the sizing of stent diameter and length~\cite{raber2018clinical}. 

Intravascular polarimetry also offers a window of opportunity for prospective identification of remote lesions with a propensity for causing subsequent acute events. 
There remains a high rate of recurrent coronary events within only a few years following initial PCI, caused in about 50\% of the cases by a lesion not involved in the original event~\cite{erlinge2021identification}. 
Plaques that rupture typically are depleted of collagen~\cite{burke1997coronary, virmani2000lessons}, and are expected to be lowly birefringent. 
The fibrous caps of target lesions in patients with chronic coronary syndrome featured indeed significantly higher birefringence than the caps of patients with acute coronary syndrome~\cite{otsuka2020intravascular}. 
Combined, intravascular polarimetry converts the polarization properties of tissues into endogenous imaging contrast that may facilitate segmentation of subsurface features and could in turn enable improved guidance of PCI, as well as refined assessment of remote lesions. 

We build on the success of deep learning methods for segmentation tasks. 
This paper proposes a convolutional neural network and optimizes its performance using a new multi-term loss function. 
It uses the additional image contrasts available to intravascular polarimetry by appending the three channels (conventional intensity, birefringence, depolarization) into a multichannel image to automatically analyze coronary artery images.

The multi-term loss function includes two common segmentation loss terms, i.e., weighted cross-entropy loss and generalized multi-class Dice loss. 
In addition to the common segmentation loss terms, a boundary loss term focuses on the accuracy of the model only within the pixels close to the boundaries between the anatomical layers. 
The boundary loss is suitable for problems requiring precise object boundary detection. 
Similarly, “boundary cardinality loss” penalizes the model from a topological point of view when the number of anatomical layers is different between the model's prediction and ground-truth by counting the number of boundary pixels along the radial axis. 
The boundary cardinality loss imposes a topological prior on the layered tissues.  Additionally, a feature denoted attending physician loss uses an independently-trained critique model, which distinguishes between low- and high-quality labels. 
The attending physician loss enables the utilization of the auxiliary information embedded in datasets with heterogeneous manual labeling qualities.

\begin{table}
\centering
\caption{The definition of the six exclusive labels that are based on the manual expert annotations and shown in Fig.~\ref{fig:annotations}.D.}
\begin{tabular}{|c|l|l|}
\hline
\textbf{Class } & \textbf{Label } & \textbf{Definition} \bigstrut\\
\hline
1     &  Outside  & external elastic lamina and deeper tissues  \bigstrut\\
\hline
2     &  Lumen  & Interior of the convex vessel lumen \bigstrut\\
\hline
3     &  Intima  & visible tunica intima  \bigstrut\\
\hline
4     &  Media  & visible tunica media  \bigstrut\\
\hline
5     &  G-Shadow & guidewire shadows between lumen and EEL  \bigstrut\\
\hline
6     &  P-Shadow & plaque shadows between lumen and EEL \bigstrut\\
\hline
\end{tabular}%

\label{tab:class_label_def}
\end{table}

\begin{figure*}
\centerline{\includegraphics[width=.95\textwidth]{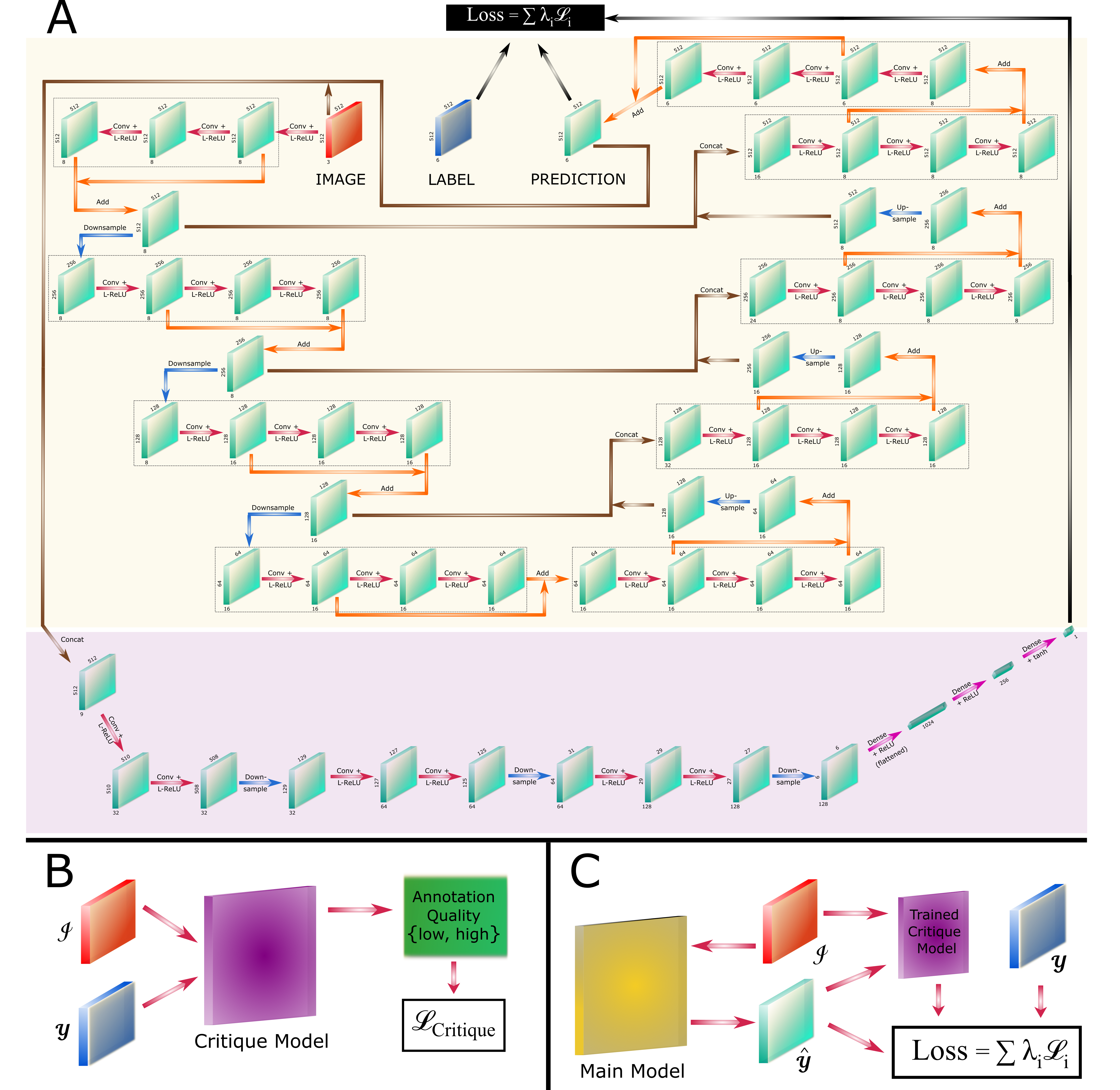}}
\caption{The proposed model architecture. The main model takes the multi-channel polarimetric image as the input and produces a multi-class probability prediction as the output ($\hat{\boldsymbol{\mathcal{Y}}}$).
The auxiliary critique model (a.k.a. Attending Physician) is trained independently with concatenated images ($\boldsymbol{\mathcal{I}}$) and ground-truth labels ($\boldsymbol{\mathcal{Y}}$) as its input to predict the quality level of manual labeling. 
This critique model then evaluates the segmentation during the main model’s training by providing one of the loss terms.}
\label{fig:architecture}
\end{figure*}

We trained and evaluated our method on a set of 984 images from 57 patients and compared it to the performance of state-of-the-art algorithms reported in the literature. Our work is the first demonstration of automatic segmentation of anatomical layers and the shadow artifacts arising from the guidewire and lipid-rich lesions using intravascular PS-OCT. It improves the boundary detection of the coronary lumen compared to other methods, and identifies the guidewire and plaque shadows in a single step. Furthermore, to the best of our knowledge, only two other studies~\cite{zahnd2017contour, chen2018quantitative} reported on the detection accuracy of intima-media and the media-adventitia boundaries, and only in regions of minimal disease, specifically excluding areas with thickened intima characteristic of atherosclerotic lesions.

\section{Related Work} \label{sec:related_works} 
The clear need for simplified interpretation of intravascular OCT images has motivated the development of automated methods, with a focus on lumen segmentation, that tackle the challenges and artifacts presented by typical intravascular OCT data.

OCT A-lines are independently acquired along the radial direction of the vessel in cylindrical coordinates as the probe rotates. 
Simultaneously, the core of the catheter is pulled back through the vessel, resulting in a helical scan pattern. 
Contemporary IV-OCT reconstructs individual A-lines from measurements in the frequency domain and visualizes the logarithm of the power of the reconstructed signal (dB). 
The presence of the guidewire, used to safely deploy the intravascular imaging catheter, casts a shadow on the vessel wall. 
The resulting intrinsic discontinuity even in the lumen signal creates an artifact that needs to be addressed by all analysis approaches. 
Prior signal processing techniques aimed to analyze A-lines independently in order to exploit the rich characteristics of the OCT signal~\cite{moraes2013automatic,bologna2019automatic,macedo2015bifurcation,macedo2016robust,akbar2019automated}. 

Researchers proposed various classical segmentation methods to detect the vessel lumen in IV-OCT images. 
The region-based active contour segmentation methods with level-set energy functions utilize the prior cross-sectional information~\cite{ughi2012automatic, joseph2016automatic, cao2017automatic}.
While dynamic programming~\cite{zhu2021automatic, cao2017automatic} or small artificial neural networks~\cite{nam2016automated} can be used to correct the remaining artifacts after the application of the primary methods, level-set methods perform poorly in low signal-to-noise (SNR) regions and produce non-smooth and imprecise lumen boundaries.
Graphical models, such as graph-cuts, have also been used for segmentation of OCT images as the various boundaries do not intersect making the graphical models well suited for this problem~\cite{tsantis2012automatic,wang2014fully}. 
This approach has been used for the detection of internal anatomical layers, i.e., the inner and outer boundary of the media ~\cite{chen2018quantitative}, or also including the outer adventitia boundary ~\cite{zahnd2017contour}, in areas of minimal intimal thickness.
The counterpart physics-based methods formulate the segmentation as a diffusion problem~\cite{roy2015lumen,olender2018mechanical,yang2020novel}. 
Despite the success of graphical methods, difficulties arise in low SNR regions with speckle, anatomical anomalies, and external objects, which limit the practical applications of these models and cause a cascade of increasing errors that require follow-up manual corrections done by expert annotators in the post-processing stage~\cite{pazdernik2018early}. 

Recently, deep learning models have emerged as a solution to many medical image analysis problems, including IV-OCT. 
Yong et al.~\cite{yong2017linear} used a regression deep learning network to detect the vessel lumen along the radial direction in polar coordinates. 
Gharaibeh et al.~\cite{gharaibeh2019coronary} segmented vessel lumen and coronary calcifications in IV-OCT images using a U-Net architecture and post-processed the output with a conditional random field model. 
Abdolmanafi et al.~\cite{abdolmanafi2018characterization, abdolmanafi2019fully} used pre-trained convolutional neural networks (CNN) to identify and classify several tissue types encountered in the coronary arteries using transfer learning.
Specific attention has also been paid to the segmentation of stent struts to confirm stent placement and identify malapposition~\cite{tsantis2012automatic,ughi2012automatic,attizzani2014mechanisms,wu2020automatic,yang2020novel}.
Our model builds on these previous approaches and extends the state-of-the-art deep learning methods to detect guidewire and plaque shadows as well as anatomical layers not only in minimally diseased vessels but also through thickened intima in coronary arteries imaged with intravascular polarimetry. 

\section{Methods} \label{sec:methods} 
This section introduces the loss functions tailored for vessel layer and artifact segmentation, describes the training and evaluation procedures, and provides implementation details.  
Fig.~\ref{fig:architecture} illustrates the architecture of our model.
As described below, one of our datasets was revised extensively to provide a curated set of high-quality segmentations. 
This practice resembles the scenario of an attending physician reviewing and scoring manual annotations performed by resident physicians to provide constructive feedback for training. 
By analogy, we trained a model to critique the multi-class labels conditioned on their input images by distinguishing between the initial and final revisions of the labels. 
The trained model was used as the attending physician loss term ($\mathcal{L}_{AP}$) to critique the quality of predicted labels by the main model.

\subsection{Segmentation Loss}
We developed a multi-term multivariate loss function that includes novel loss terms. 
The first loss term is the weighted cross-entropy function 
\begin{align}
\mathcal{L}_{WCE} (\mathbf{y}, \hat{\mathbf{y}}) = - \frac{1}{N} \sum_{c=1}^{N_c} \omega_c \sum_{i,j} \left[ \mathbf{y}_{ijc} \log \hat{\mathbf{y}}_{ijc} \right] 
\end{align}
that measures the cross-entropy between the target label $y$ and predicted label probabilities $\hat{y}$ of all $N$ pixels where $i$ and $j$ are 2D matrix indices,  $c$ is the class index, and $N_c$ is the number of classes. 
Each pixel's cross-entropy is then weighted proportionately to the inverse of its class population ($\omega_c=\lVert \mathbf{Y} \rVert_1 / \lVert \mathbf{Y}_c \rVert_1$). 

The second loss term is a multi-class version of the generalized Dice loss function~\cite{milletari2016v}
\begin{align}
\mathcal{L}_{Dice} (\mathbf{y}, \hat{\mathbf{y}}) = 1 - \frac{2}{N_c} \sum_{c=1}^{N_c} \frac{\sum_{i,j} \left[ \mathbf{y}_{ijc} \times \hat{\mathbf{y}}_{ijc} \right] + \epsilon}{\sum_{i,j} \left[\mathbf{y}_{ijc}^2 + \hat{\mathbf{y}}_{ijc}^2 \right] + \epsilon}
\label{eq_dice_loss}
\end{align}
that measures each label's segmentation accuracy similar to the Dice coefficient.
Dice loss uses the prediction probability (e.g., softmax of logits) instead of the classification result and ranges from zero to one, with zero corresponding to the most accurate result.
We add a small constant ($\epsilon$) for numerical stability.
These segmentation loss functions are used widely in the field~\cite{milletari2016v,litjens2017survey}.

\subsection{Boundary Loss}
The weighted cross-entropy and Dice loss terms are only marginally affected by the errors on the boundary because the boundary pixels are a small portion of the target objects. 
The third loss term is based on the boundary segmentation accuracy and focuses the network’s attention on the close vicinity of label boundaries. 
The boundary precision loss term ($\mathcal{L}_{BP}$) utilizes a boundary neighborhood mask,
\begin{align}
\begin{split}
\pmb{\beta} (\mathbf{y}) =&
\left( \bigvee_{c=1}^{N_c} \left[ \left( \mathbf{1}_{b \times 1} \ast \mathbf{y}_c \right) \oplus \left( \mathbf{y}_c \right) \right] \right) \bigvee \\
&\left( \bigvee_{c=1}^{N_c} \left[ \left( \mathbf{1}_{b \times 1} \ast (1 - \mathbf{y}_c) \right) \oplus \left( 1 - \mathbf{y}_c \right) \right] \right), 
\end{split}  
\end{align}
which masks the cross-entropy loss values of the pixels that are not in close vicinity of the label boundaries along the radial axis (e.g. $> b$, $b=10$ pixels) using all-ones matrix ($\mathbf{1}$) and two logical operators, i.e., convolution ($\ast$), disjunction (${\scriptstyle  \bigvee}$) and exclusive disjunction ($\oplus$). The boundary precision loss term, 
\begin{align}
\mathcal{L}_{BP} (\mathbf{y}, \hat{\mathbf{y}}, \pmb{\beta}) =& 
\frac{-1}{\sum_{i,j} \pmb{\beta}_{ij}} \sum_{c=1}^{N_c} \sum_{i,j} \pmb{\beta}_{ij} \mathbf{y}_{ijc} \log \hat{\mathbf{y}}_{ijc},
\end{align} 
is differentiable with respect to the model parameters as long as $\pmb{\beta}$ is a function of the ground truth target label. 

\subsection{Attending Physician Loss}
The training of a critique model involves a loss function that measures the distance of model parameters from the optimal solution in the parameter space. 
Arjovsky et al.~\cite{arjovsky2017wasserstein} proposed \textit{Wasserstein-1} (a.k.a. Earth-Mover) distance, 
\begin{align}
W_1(P,P') = \inf_{\gamma \in \Pi(P,P')} \mathbb{E}_{(x,x') \sim \gamma} \lVert x - x' \rVert_1 , 
\label{eq.wasserstein1}
\end{align}
where $\Pi(P,P')$ is the set of all joint distributions $\gamma(x,x')$ that their marginal distributions are equal to $P$ and $P'$. 
Wasserstein-1 is the optimal cost of transporting a mass with distribution $P$ to another mass with distribution $P'$ when the transport cost and transport distance are linearly related.
Stable learning with a meaningful learning curve that avoids common problems, including mode collapse, can be obtained when \textit{Wasserstein-1} distance is adopted~\cite{arjovsky2017wasserstein}. 
Since the infimum in (\ref{eq.wasserstein1}) is intractable, Kantorovich and Rubinstein~\cite{villani2008optimal} proposed a tractable dual problem, 
\begin{align}
W_1(P,P') = \sup_{\lVert f \rVert_L \leq 1} \{
\mathbb{E}_{x \sim P} [f(x)] - \mathbb{E}_{x' \sim P'} [f(x')] \}, 
\end{align}
where $f$ is a 1-Lipschitz function, mapping the support of $P$ and $P'$ to real numbers. 

Similarly, we can select $f$ from a family of parameterized functions ($\{f_w\}_{w \in \mathcal{W}}$) that are at least K-Lipschitz for a constant K and optimize (\ref{eq.wasserstein1}) over the functional parameter space, 
\begin{align}
W_1(P,P') &= \max_{w \in \mathcal{W}} 
\{\mathbb{E}_{x \sim P} [f_w(x)] - \mathbb{E}_{x' \sim P'} [f_w(x')]\}, 
\end{align}

The requirement of $f$ being K-Lipschitz for the function family of deep neural networks can be imposed by clipping the parameter values with an absolute value upper limit \cite{arjovsky2017wasserstein} or enforcing the gradient of parameters to be 1 almost everywhere through a gradient penalty loss term \cite{gulrajani2017improved}. 
Gulrajani et al.~\cite{gulrajani2017improved} showed that the gradient constraining method improves the learning process compared to the weight clipping method. 
We observed a similar effect while training our Attending Physician model, which is then used as the fourth loss term 
\begin{align}
\mathcal{L}_{AP} = \mathbb{E}_{x \sim P} [f_w(x)] - \mathbb{E}_{x' \sim P'} [f_w(x')]. 
\end{align}
Therefore, we trained the main model from both types of manual segmentations by introducing the critique model (Fig.~\ref{fig:architecture}).

\subsection{Topological Loss}
The last loss term examines the labels from a topological point of view. 
Ideally, the predicted labels along A-lines are composed of three or four connected components without any void, starting with the lumen label in the center and ending with the outside label. 
The area between the lumen and outside labels should be occupied by two adjacent solid anatomical layers (i.e., intima and media) or by one of the artifact labels (i.e., guidewire or plaque shadows). 
These configurations are distinguishable in terms of the number of label boundaries along the radial direction. 
The soft boundary cardinality loss term ($\mathcal{L}_{BC}$) penalizes the discrepancy between the predicted and ground truth labels based on the number of boundary pixels along the radial axis. 
We propose to employ
\begin{align}
\mathbf{S} (\mathbf{y}) = 1 + \tanh{\left( M \left[ \mathbf{y} - \max_{c} \mathbf{y}_{ijc} \right] \right)}
\end{align}
that is a differentiable proxy for the arguments of the maxima and a saturated equivalent of the softmax function, i.e., $e^\mathbf{x}/ \lVert e^\mathbf{x_i} \rVert_1$. 
The soft argmax admits the predicted class probabilities at each pixel and maps the probability of the most probable class and other classes to $\sim 1$ and $\sim 0$, respectively. 
The level of saturation is controlled by the large number $M$ and the precision of the probability values. 
Since the value of the soft argmax for a given class changes between two adjacent pixels at the label boundary, the soft boundary set cardinality along the radial axis
\begin{align}
\mathbf{BC} (\mathbf{S}) = \frac{1}{2} \sum_i \sum_c \lvert \mathbf{S}(\mathbf{y})_{(i+1)jc} - \mathbf{S}(\mathbf{y})_{ijc} \rvert
\end{align}
approximates the number of class boundaries in each A-line.
The boundary cardinality loss function  
\begin{align}
\mathcal{L}_{BC} (\mathbf{y}, \hat{\mathbf{y}}) = \sigma(\mathbf{BC} (\mathbf{S}(\mathbf{y})), \mathbf{BC} (\mathbf{S}(\hat{\mathbf{y}})))
\end{align}
compares the prediction and ground-truth labels with respect to the number of boundaries, where $\sigma$ measures the difference between two $\mathbf{BC}$ vectors (e.g., norm 1). 
We considered 1, 100, and $100/\epsilon$ for $M$, where $\epsilon$ is the small number used for mathematical stability in (\ref{eq_dice_loss}) and sofmax value clipping. 
For $\sigma$, we considered $\lVert \cdot \rVert_1$, $\lVert \cdot \rVert_2$, and $\max(\cdot)$. 
Based on the validation dataset and the convexity of $\mathcal{L}_{BC}$, norm-1 ($\lVert \cdot \rVert_1$) and $100/\epsilon$ are the optimal choices for $\sigma$ and $M$, respectively.

The final loss function combines all five loss terms: 
\begin{align}
\begin{split}
\mathcal{L} =& \lambda_{WCE} \mathcal{L}_{WCE} + \lambda_{Dice} \mathcal{L}_{Dice} + \\
& \lambda_{BP} \mathcal{L}_{BP} + \lambda_{AP} \mathcal{L}_{AP} + \lambda_{BC} \mathcal{L}_{BC},
\end{split}
\end{align}
in which loss term weights ($\lambda_.$) are selected within the range $[10^{-3}, 10^{3}]$ and optimized over their logarithmically-spaced multidimensional grid using greedy algorithms. 

\subsection{CNN Architecture}
The proposed network architecture scheme is based on the U-Net \cite{ronneberger2015u} and deep residual learning \cite{he2016deep} models (Fig.~\ref{fig:architecture}). 
The auxiliary critique model was trained independently to distinguish low- and high-quality labels. 
Subsequently, the main model was trained by combining the trained critique model and other loss terms to segment three anatomical layers and two shadows artifacts (Table~\ref{tab:class_label_def}). 
The optimized architecture contains multi-scale encoder and decoder sections with skip connections at each scale. 
The input consists of the three-channel images of conventional intensity, birefringence, and depolarization, in polar coordinates, down sampled to 512 by 512 pixels. The output consists of the six concatenated classes of the same pixel dimension. 
The convolutional complex contains three convolutional layers with a $3 \times 3$ pixel kernel size and a leaky version of the rectified linear unit (L-ReLU) activation function, which has a negative slope coefficient of 0.3. 
These three convolutional layers compute the residual values by using an internal skip connection. 
The max-pooling layers with a $2 \times 2$ pixel kernel size are applied after convolutional complexes in the encoding section for down-scaling while the counterpart deconvolutional layers are applied for bi-linear up-scaling within the decoding section. 
The encoding output and decoding input are connected through two convolutional complexes that operate at the latent representation level. 
The layers within each of the three scales and the latent representation layers have 8, 8, 16, and 16 features, respectively. 

The critique model architecture accepts the concatenation of image channels and output label channels as the input and applies three convolutional complexes with 32, 64, and 128 features, respectively. 
Each complex consists of two convolutional layers with a $3 \times 3$ pixel kernel size and the ReLU activation function followed by a max-pooling with a $2 \times 2$ pixel kernel size.
The last complex's output is flattened and processed by a three-layer dense neural network with 1024, 256, and 128 hidden nodes and ReLU activation function, respectively. 
The final output has one feature and uses the hyperbolic tangent activation function (Fig.~\ref{fig:architecture}). 
\subsection{Training and Implementation}
We randomly divided the annotated dataset between training, validation, and hold-out testing dataset by selecting 45, 6, and 6 patients (80\%/10\%/10\%), respectively.
Augmentation included random mirroring, rotation, multi-channel image intensity distribution manipulations ([-0.05, 0.05] brightness and [0.9, 1.1] contrast), and spatial scaling ([0.875, 112.5]). 
An element from the power set of the image augmentation set was applied to each given PS-OCT cross-section with randomly selected transformation parameters sampled uniformly and independently from the ranges above. 
The geometric transformations were defined in the Cartesian coordinate system, but they were implemented and applied in the polar coordinate system. 
The data augmentation methods were implemented and executed on a GPU to improve the model's runtime.

We implemented our model in Python using Keras\textsuperscript{\texttrademark} and Tensorflow\textsuperscript{\texttrademark}. 
We commonly used RMSprop optimizer with $10^{-3}-10^{-4}$ learning rate and mini-batch size of 20 per GPU.
The GPU memory size was the limiting factor in the learning rate and mini-batch size selection. 
We used two NVIDIA\textsuperscript{\textregistered} GeForce\textsuperscript{\textregistered} RTX 2080 Ti or four NVIDIA\textsuperscript{\textregistered} Tesla\textsuperscript{\textregistered} V100.

\subsection{Post-processing}
We investigated a post-processing procedure to the model output to enforce known topology of the multi-class segmentations. 
Initially, small objects and holes within each class were removed, and their interfaces were smoothed. 
Then, a set of logical operations was applied to impose the topological relationships between the classes in the polar coordinate system. 
The proposed set includes the following constraints:
\begin{itemize}
    \item Lumen is a single connected object without any 2D void. The same rule applies to both guidewire shadow and outside. 
    \item Guidewire and plaque shadows are confined between the lumen, the outside, and two A-lines.
    \item The order of layers from inside to outside ends is lumen, intima, media, and outside.
\end{itemize}

\subsection{Performance Metrics}
Based on the ground-truth labels, we evaluated the performance of the multi-class prediction model using accuracy and Dice coefficient, 
\begin{align}
Dice(\mathbf{Y}, \Hat{\mathbf{Y}}) &= \frac{2}{\lVert \mathcal{C} \rVert}\sum_{c\in \mathcal{C}} \frac{\lVert \mathbf{Y}_c \cap \Hat{\mathbf{Y}}_c\rVert}{\lVert \mathbf{Y}_c \rVert + \lVert {\Hat{\mathbf{Y}}}_c \rVert}, 
\end{align}
where $\lVert \cdot \rVert$ is the set cardinality, $\Hat{\mathbf{Y}}_c$ is the set of predicted pixels as class $c$, $\mathbf{Y}_c$ is the set of pixels in ground-truth as class $c$, $\mathcal{C}$ is the set of classes, and $\cap$ is the intersection operation. 
Furthermore, we evaluated the precision of inter-class boundaries using the average distance error (ADE) along the radial direction and modified Hausdorff distance (MHD) \cite{dubuisson1994modified} in 2D within the cross-section:
\begin{align}
d(a, \mathcal{B}) &= \inf_{b \in \mathcal{B}} \{ \lVert a - b \rVert_2 \}, \\
ADE(\Hat{\mathcal{B}}; \mathcal{B}) &= \frac{1}{\lVert \Hat{\mathcal{B}} \rVert} \sum_{a \in \Hat{\mathcal{B}}} d(a, \mathcal{B}), \\
MHD(\Hat{\mathcal{B}}, \mathcal{B}) &= \max \{ ADE(\Hat{\mathcal{B}}; \mathcal{B}) , ADE(\mathcal{B}; \Hat{\mathcal{B}}) \}, 
\end{align}
where $\mathcal{B}$ and $\Hat{\mathcal{B}}$ are the set of boundary pixels in the ground truth and prediction, respectively, and $\lVert . \rVert_2$ is the Euclidean norm. 

\section{Data}
We demonstrate the method on images from an intravascular polarimetry pilot study, which included two cohorts and enrolled a total of 57 patients who underwent percutaneous coronary intervention and PS-OCT imaging at the Erasmus University Medical Center in Rotterdam. 
Of the 57 pullbacks, only segments of native vessel wall or containing old stents from previous interventions were included in this study. 
The Ethics Committee of Erasmus Medical Center approved the study protocol, and all procedures were performed in accordance with local and federal regulations and the Declaration of Helsinki.  

The imaging system consists of ``FastView'' intravascular catheters (Terumo Co., Tokyo, Japan) interfaced with our custom-built PS-OCT system, operating at 1300 nm central wavelength similar to commercially available clinical IV-OCT systems. 
The wavelength scanning range was 110 nm, achieving a radial resolution below 10 $\mu$m, assuming a tissue refractive index of 1.34. The dimension of the pixels in the reconstructed tomograms in the radial direction were 4.2 $\mu$m and 4.43 $\mu$m, respectively, for the two cohorts. 
The catheter's rotation speed was 100 RPS, with 1024 radial scans per rotation, and pullbacks were performed at 10 mm/s or 20 mm/s, at the operator’s discretion. 
Non-ionic contrast solution was injected at a rate of 3-4 mL/s during the pullback to displace coronary blood and obtain an unperturbed view of the vessel wall. 

Intravascular polarimetry was performed based on our earlier work~\cite{villiger2018coronary,otsuka2019intravascular, otsuka2020intravascular,otsuka2020polarimetric,otsuka2020intravascularB}. 
Briefly, an electro-optic polarization modulator was used to alternate the polarization state of the light incident on the tissue between consecutive depth scans and a polarization-diverse receiver enabled determination of the detected light's polarization state and intensity. 
Polarimetric analysis employed spectral binning ~\cite{villiger2013spectral} to reconstruct maps of tissue birefringence and depolarization. 
Birefringence is the difference in the refractive index experienced by orthogonal polarization states aligned and orthogonal to the tissue optic axis, respectively. 
Tissue depolarization measures the randomness of the detected light’s polarization state using the complement to one of the degree of polarization. 

Initially, an expert interventional cardiologist (K.O.) excluded partial segments of 3D pullbacks that were uninterpretable and suffered from severe artifacts caused by insufficient blood clearing. 
The qualified pullback segments added up to 3936 mm of pullbacks at a 100 or 200 $\mu$m pitch. 
Subsequently, the expert annotated a total of 984 PS-OCT cross-sections spaced 4 mm apart using our in-house Matlab graphical user interface (Fig.~\ref{fig:annotations}.D), using the conventional OCT signal as well as the polarization channels. 
The manual annotations included the outer boundaries of the lumen, tunica intima (i.e., internal elastic lamina (IEL)), and tunica media (i.e., external elastic lamina (EEL)).
The location of IEL and EEL within the plaque and guidewire shadows were extrapolated based on their visible segments (Fig.~\ref{fig:annotations}.E). 
Additionally, angular segments containing plaque, guidewire, stent struts, side branches, or thrombus were identified and used for segmentation or selective analysis without influencing the main label categories.
Consequently, as summarized in Table~\ref{tab:class_label_def}, the manual annotations were converted into six exclusive labels: outside, lumen, visible intima, visible media, plaque shadow, and guidewire shadow (Fig.~\ref{fig:annotations}.F). 

To manage the workload, we annotated the total dataset in four separate batches and through three phases: initial annotation, high-precision annotation, and annotation approval. 
One of the batches was revised extensively at the pixel-level, requiring four times as long as other batches. 
The high-accuracy batch, in combination with its initial annotation, was utilized to train the proposed critique model and its resulting loss term.

\section{Results} \label{sec:results}
\begin{figure*} 
\centerline{\includegraphics[width=.95\textwidth]{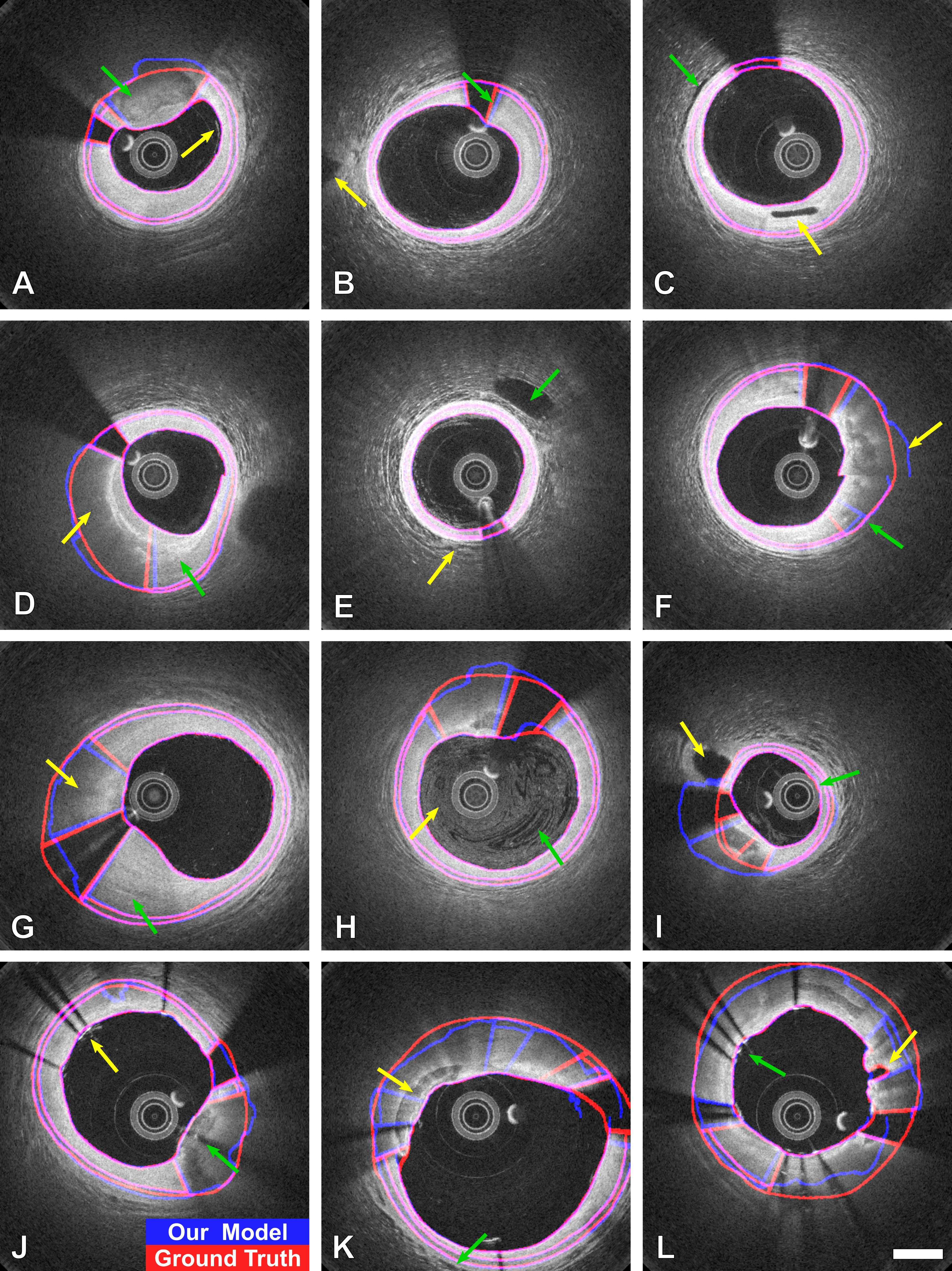}}
\caption{Qualitative assessment of PS-OCT cross-sections. 
The annotations of our model and the ground-truth are overlaid on the gray-scale intensity image in blue and red outlines, respectively. 
See the text for detailed discussion. 
Scale bar: 1 mm.}
\label{fig:qualitative_cf}
\end{figure*}

We compared the model's automated annotation results to the expert's ground-truth annotations in Fig.~\ref{fig:qualitative_cf} to qualitatively characterize our model, illustrate the model's strengths, and identify possible areas of improvement. 
Our model's annotations and the ground-truth are overlaid on the gray-scale intensity image in blue and red outlines, respectively. 

The most common complication for boundary annotation, particularly for the outer intima and outer media, is the presence of thick plaques or calcium (e.g. Fig.~\ref{fig:qualitative_cf}.A\, green arrow) and thickened vessel walls (e.g. Fig.~\ref{fig:qualitative_cf}.G, green arrow) that cause significant reduction in the detected signal.
The background signal and statistical noise characteristics within the plaque regions impede the model's objective to annotate the anatomical layers and result in higher annotation variability (e.g. Fig.~\ref{fig:qualitative_cf}.F, yellow arrow; Fig.~\ref{fig:qualitative_cf}.G, yellow arrow; Fig.~\ref{fig:qualitative_cf}.K, yellow arrow). 

Nonetheless, whenever the image information supports the ground-truth boundaries, the model matches well with the expert annotations even in these challenging cases (e.g. Fig.~\ref{fig:qualitative_cf}.D, both arrows). 
Correspondingly, the boundaries detected by the model may conform with the underlying multi-dimensional images more accurately than the ground-truth annotations (e.g. Fig.~\ref{fig:qualitative_cf}.F, green arrow), suggesting inconsistencies in the manual ground-truth segmentation.

The guidewire obstructs the probing light, causing a fuzzy signal at its boundaries, resulting in imprecise automatic and manual boundary detection (e.g. Fig.~\ref{fig:qualitative_cf}.B, green arrow). 
Moreover, the physical proximity of the vessel lumen with the guidewire and catheter leads to perturbed pixel-level delineation of the lumen boundary (e.g. Fig.~\ref{fig:qualitative_cf}.B, green arrow; Fig.~\ref{fig:qualitative_cf}.I, green arrow). 

Side branches can appear in various locations of the field of view and could be expected to exhibit confusing features, yet our model analyzes these cases in concordance with the ground truth annotation. 
Such vessels might appear outside the vessel wall (e.g. Fig.~\ref{fig:qualitative_cf}.B, yellow arrow; Fig.~\ref{fig:qualitative_cf}.E, green arrow), directly adjacent to the vessel wall boundary (e.g. Fig.~\ref{fig:qualitative_cf}.I, yellow arrow), inside the intima (e.g. Fig.~\ref{fig:qualitative_cf}.C, yellow arrow), or in direct communication with the lumen (e.g. Fig.~\ref{fig:qualitative_cf}.L, yellow arrow).

Even though non-ionic contrast solution is injected during catheter pull-back to displace blood, residues of blood may persist in the vessel lumen vicinity (e.g. Fig.~\ref{fig:qualitative_cf}.A, yellow arrow; Fig.~\ref{fig:qualitative_cf}.E, yellow arrow). 
Blood clearance can be incomplete, especially at the onset or the end of contrast injection Fig.~\ref{fig:qualitative_cf}.H, both arrows). 
Still, in all these cases, our model successfully detects the lumen outer boundaries. 
Equivalently, the dark and bright tissue patterns (e.g. Fig.~\ref{fig:qualitative_cf}.C, green arrow; and Fig.~\ref{fig:qualitative_cf}.K, green arrow) are observed beyond the media layer and mimic the multi-layer vessel wall structures but they do not distract the automatic boundary allocations. 

While our study only included intravascular imaging prior to intervention, previously embedded stents are commonly encountered, owing to the high recurrence rate of acute coronary syndrome and myocardial infarction. Depending on the specific stent material and patient history, stents might appear embedded in the vessel wall (e.g. Fig.~\ref{fig:qualitative_cf}.J, green arrow) or protruding into the lumen (e.g. Fig.~\ref{fig:qualitative_cf}.J, yellow arrow; Fig.~\ref{fig:qualitative_cf}.L, green arrow).
Stents generate diverse and strong image artifacts that impede the model's ability to correctly detect the boundaries. 
Exact layer segmentation behind stents presents challenges even for expert readers. With the exception of neointimal hyperplasia, previously stented segments are unlikely to reside in the culprit segment. Such segments were included in our data set merely to train the model to ignore the ensuing artifacts. Notably, there exists a distinct class of models designed to detect stent struts and verify correct stent deployment~\cite{tsantis2012automatic,ughi2012automatic,yang2020novel}). 

To complement the qualitative assessment of the model with quantitative metrics Table~\ref{tab:class_accuracy} lists the model’s performance for individual label classes, evaluated on the hold-out test set. The lumen segmentation achieved the best scores for all metrics while the plaque shadow performance influenced by the more ambiguous ground-truth labels owing to the lack of clear structural markers. Nonetheless, the individual metrics confirm the overall high quality of segmentation achieved by the model.

\begin{table}
\centering
\caption{The performance of our multi-label classification model based on different performance metrics.}
\begin{tabular}{|l|c|c|c|c|}
\hline
\textbf{Label } & \textbf{Sensitivity} & \textbf{Specificity} & \textbf{Accuracy} & \textbf{Dice} \bigstrut\\
\hline
 Outside  & 99.0\% & 99.5\% & 99.3\% & 99.3\% \bigstrut\\
\hline
 Lumen  & 99.7\% & 99.9\% & 99.8\% & 99.7\% \bigstrut\\
\hline
 Intima  & 86.1\% & 99.7\% & 98.4\% & 91.0\% \bigstrut\\
\hline
 Media  & 78.2\% & 99.5\% & 99.0\% & 79.5\% \bigstrut\\
\hline
 G-Shadow & 94.6\% & 97.5\% & 97.3\% & 83.8\% \bigstrut\\
\hline
 P-Shadow & 86.7\% & 80.2\% & 82.4\% & 76.7\% \bigstrut\\
\hline
\end{tabular}%

\label{tab:class_accuracy}
\end{table}

\begin{table}
\centering
\caption{Loss terms ablation study.}
\begin{tabular}{|c|c|c|c|c|c|c|c|}
\hline
\begin{sideways}$\mathcal{L}_{WCE}$\end{sideways} & \begin{sideways}$\mathcal{L}_{Dice}$\end{sideways} & \begin{sideways}$\mathcal{L}_{BP}$\end{sideways} & \begin{sideways}$\mathcal{L}_{AP}$\end{sideways} & \begin{sideways}$\mathcal{L}_{BC}$\end{sideways} & \textbf{Accuracy} & \textbf{Dice} & \textbf{MHD} \bigstrut\\
\hline
$\checkmark$ & $\checkmark$ & $\checkmark$ & $\checkmark$ & $\checkmark$ & 96.0\% & 88.3\% & 2.3 \bigstrut\\
\hline
$\checkmark$ & $\checkmark$ & $\checkmark$ & $\checkmark$ & $\times$ & 95.3\% & 86.5\% & 3.2 \bigstrut\\
\hline
$\checkmark$ & $\checkmark$ & $\checkmark$ & $\times$ & $\times$ & 96.1\% & 88.3\% & 3.8 \bigstrut\\
\hline
$\checkmark$ & $\checkmark$ & $\times$ & $\times$ & $\times$ & 94.3\% & 84.7\% & 6.0 \bigstrut\\
\hline
$\checkmark$ & $\times$ & $\times$ & $\times$ & $\times$ & 94.6\% & 83.8\% & 7.2 \bigstrut\\
\hline
\end{tabular}%

\label{tab:loss_term_results}
\end{table}

\begin{table}
\centering
\caption{Comparison of our model and other studies that segment the vessel lumen.}
\begin{tabular}{|l|c|c|}
\hline
\textbf{Model } & \textbf{\makecell{Dice \\ Coefficient}} & \textbf{\makecell{Other \\ Metrics}} \bigstrut\\
\hline
Abdolmanafi et al. \cite{abdolmanafi2019fully} & -     & Acc=96\% \bigstrut\\
\hline
Akbar et al. \cite{akbar2019automated} & -     & TPR=93.1\% \bigstrut\\
\hline
Bologna et al. \cite{bologna2019automatic} & -     & \makecell{TPR=97.4\% \\ TNR=99.5\%} \bigstrut\\
\hline
Cao et al. \cite{cao2017automatic} & 98.1\% & - \bigstrut\\
\hline
Cheimariotis et al. \cite{cheimariotis2017arcoct} & 93.5\% & - \bigstrut\\
\hline
Chen et al. \cite{chen2018quantitative} & -     & ADE=2.37 $\mu m$ \bigstrut\\
\hline
Gharaibeh et al. \cite{gharaibeh2019coronary} & 98\%  & Acc=98\% \bigstrut\\
\hline
Joseph et al. \cite{joseph2016automatic} & 78\%  & - \bigstrut\\
\hline
Macedo et al. \cite{macedo2015bifurcation} & 97.5\% & - \bigstrut\\
\hline
Macedo et al. \cite{macedo2016robust} & 97.0\% & - \bigstrut\\
\hline
Moraes et al. \cite{moraes2013automatic} & 97.1\% & - \bigstrut\\
\hline
Olender et al. \cite{olender2018mechanical} & 95.9\% & - \bigstrut\\
\hline
Olender et al. \cite{olender2019simultaneous} & -     & Acc=94.9\% \bigstrut\\
\hline
Tsantis et al. \cite{tsantis2012automatic} & 96.7\% & - \bigstrut\\
\hline
Tung et al. \cite{tung2011automatical} & 97\%  & - \bigstrut\\
\hline
Ughi et al. \cite{ughi2012automatic,cao2017automatic} & 96.9\% & - \bigstrut\\
\hline
Wang et al. \cite{wang2011automatic} & 97\%  & - \bigstrut\\
\hline
Yang et al. \cite{yang2020novel} & 97.6\% & - \bigstrut\\
\hline
Yong et al. \cite{yong2017linear} & 98.5\% & - \bigstrut\\
\hline
Zhu et al. \cite{zhu2021automatic} & 94.6\% & Acc=98.0\% \bigstrut\\
\hline
\\\textbf{Our method} & \textbf{99.7\%} & \boldmath{}\textbf{\makecell{ADE=2.36 $\mu m$ \\ Acc=99.8\% \\ TPR=99.7\% \\ TNR=99.9\%}}\unboldmath{} \bigstrut\\
\hline
\end{tabular}%

\label{tab:lumen_cf_result}
\end{table}

To substantiate the design of the model, we conducted an ablation study to examine the individual effects of the various loss terms, i.e., 
\begin{enumerate}
    \item The soft boundary cardinality loss term ($\mathcal{L}_{BC}$).
    \item The Attending Physician (a.k.a. Wasserstein critique model) loss term ($\mathcal{L}_{AP}$), 
    \item The boundary precision loss term ($\mathcal{L}_{BP}$),
    \item The generalized soft multi-class dice loss term ($\mathcal{L}_{Dice}$), and
    \item The weighted cross-entropy loss term ($\mathcal{L}_{WCE}$),
\end{enumerate}
We measured the accuracy, Dice coefficient, and modified Hausdorff distance (MHD) averaged among all label classes on the hold-out test dataset for models trained with a reduced number of loss terms.
The results of the ablation study on the loss terms are tabulated in Table~\ref{tab:loss_term_results} and confirm that each loss term contributes to the performance of the method or the output quality. 

Before developing and refining the individual loss terms, we set out to confirm the advantage of using intravascular polarimetry compared to the conventional IV-OCT for the visualization and segmentation of anatomical layers. Using the proposed architecture we trained the model with only the weighted cross-entropy loss function ($\mathcal{L}_{WCE}$) and compared its performance to an adapted model that was trained with only the single intensity channel as input. The Dice coefficient of the media class using intravascular polarimetry data was 70.7\%, while it was only 62.7/\% when using conventional IV-OCT. The subsequent optimization of the model’s performance improved the Dice coefficient of the media class using PS-OCT to 79.5\%. The significant gain in performance achieved by using the polarimetric channels even with only the $\mathcal{L}_{WCE}$ loss term confirms our previous qualitative observations of improved contrast for the media layer ~\cite{villiger2018coronary,villiger2018repeatability}.

\begin{table}
\centering
\caption{Comparison of our model and other studies that detect the outer boundary of lumen, intima, and media by reporting mean $\pm$ standard deviation.}
% Table generated by Excel2LaTeX from sheet 'tab_boundary_result'
\begin{tabular}{|l|c|c|c|}
\hline
\multirow{2}[4]{*}{\textbf{Study}} & \multicolumn{3}{c|}{\boldmath{}\textbf{Absolute Distance Error ($\mu$m)}\unboldmath{}} \bigstrut\\
\cline{2-4}      & \textbf{Outer Lumen} & \textbf{Outer Intima} & \textbf{Outer Media} \bigstrut\\
\hline
Zahnd et al. \cite{zahnd2017contour} & -     & $29 \pm 46$ & $30 \pm 50$ \bigstrut\\
\hline
Chen et al. \cite{chen2018quantitative} & $2.37 \pm 1.84$ & $13.61 \pm 27.22$ & $16.43 \pm 30.32$ \bigstrut\\
\hline
Our method & $2.36 \pm 3.88$ & $6.89 \pm 9.99$ & $7.53 \pm 8.64$ \bigstrut\\
\hline
\end{tabular}%

\label{tab:boundary_result_cf}
\end{table}

For comparison with previous segmentation efforts using conventional IV-OCT, we compiled the reported performance of previous studies that developed segmentation methods for the lumen in Table \ref{tab:lumen_cf_result} and for those that segmented the two additional anatomical layers in Table \ref{tab:boundary_result_cf}. There are many methods (\cite{ughi2012automatic,cao2017automatic,tung2011automatical,wang2011automatic,moraes2013automatic,bologna2019automatic,macedo2015bifurcation,tsantis2012automatic,yong2017linear,macedo2016robust,olender2018mechanical,chen2018quantitative,yang2020novel,akbar2019automated,abdolmanafi2019fully,gharaibeh2019coronary,olender2019simultaneous}) that extract the lumen with Dice 95-95\% and our method outperforms them all at 99\%. Moreover, Table~\ref{tab:boundary_result_cf} indicates that our model achieves lower absolute distance error (ADE) for both outer intima and media boundaries compared to the two other reports accomplishing and reporting on this task. Here, we excluded thickened vessel walls from evaluation of the outer boundaries in Table~\ref{tab:boundary_result_cf}, in line with the analysis in ~\cite{zahnd2017contour}, which only evaluated layer segmentation in ‘healthy regions’, and ~\cite{chen2018quantitative}, which inspected allograft vessels with minimal intimal thickening. However, thickened vessel walls are the result of coronary atherosclerosis and very common, especially in the population of patients likely to undergo intravascular imaging. Importantly, our model is able to segment cross-sections including thickened vessel wall segments, although imaging through this additional tissue degrades the achieved ADE (2.60, 16.9, and 20.85 $\mu m$ for outer the lumen, intima, and media, respectively). Still, these values are comparable to the previous methods that detect these layers only in segments with minimal disease.

\section{Discussion} \label{sec:discussion}
PS-OCT complements the IV-OCT backscatter intensity signal by measuring the polarization state of reflected light and reconstructing tissue birefringence and depolarization signals. These polarimetric signals provide a more detailed characterization of the vessel wall and can help to differentiate tissue layers that have comparable scattering properties but distinct polarization features. PS-OCT enriches the visualization of anatomical layers and hence facilitates downstream image processing tasks. 
We proposed a convolutional neural network model with a new multi-term loss function that leverages the increased contrast available to PS-OCT to segment the vessel lumen, as well as the intima-media and the media-adventitia boundary.
Furthermore, the model works on all plaque types and correctly segments the inner and outer media boundaries even through thickened vessel walls, as long as the plaque is not opaque. 
Conversely, angular segments of lipid-rich or calcified plaques that impede detection of the subluminal anatomical layers are identified as plaque shadows. The model, however, continues to estimate the outer media boundary throughout these opaque regions. The model also identifies guidewire shadows without interrupting the lumen and out media segmentation.

Our comprehensive multi-class image segmentation model can support many downstream image analysis tasks. 
Automated and objective image segmentation simplifies both clinical research and affords integration into the clinical workflow by removing the workload of manual segmentation. For guidance of PCI, robust and automated measurement of the EEL diameter would simplify stent sizing~\cite{ali2016optical}. Evaluation of the intimal thickness along the vessel could enhance the common simplified visualization of the culprit vessel based on the lumen diameter with complementary information on the location and extent of plaques to select a suitable landing zone. In a clinical research setting, automated segmentation of the intimal thickness along entire coronary vessels would enable the formulation of questions that are currently impractical to address due to the workload of manual segmentation. Crucially, automated segmentation also enables evaluation of tissue polarization properties in distinct anatomical areas, which previously relied on tedious manual segmentation~\cite{villiger2018coronary,otsuka2019intravascular, otsuka2020intravascular}. We anticipate that such volumetric analysis of polarization properties will offer refined insight into plaque composition and may enable the formulation of polarization-informed plaque index similar to the lipid-core burden index of near infrared spectroscopy~\cite{erlinge2021identification}. 

The high-performance segmentation of the lumen and outside classes is an indication that we approached the limits imposed by using a single-reader ground-truth. The media boundaries and shadow classes likely suffer from higher intra-reader ground-truth variability. The anatomical layers beyond the lumen are located in areas of decaying signal quality and the shadows intrinsically have a poorly defined border. The increased boundary to area ratio of the media furthermore deteriorates typical segmentation metrics even without degradation in the boundary precision.

In addition to the use of a single-reader ground truth, the limitations of the current study include a modest number of pullbacks and a limited spectrum of atherosclerotic disease. Also, segmentation was performed on individual cross-sections. While segmentation of adjacent cross-sections enables volumetric segmentation, there may be information embedded in the volumetric data that escapes the current model. Lastly, intravascular polarimetry with PS-OCT uses commercial clinical imaging catheters but currently uses a custom imaging console, which complicates clinical translation. Towards resolving this limitation, Xiong et al.~\cite{xiong2019constrained} proposed a new method that may be compatible with existing imaging consoles and could accelerate the clinical translation of using anatomical layer segmentation based on tissue polarization properties. 

\section{Conclusion} \label{sec:conclusion}
We proposed a method for the segmentation of intravascular polarimetry images of coronary arteries. The method performance compares favorably with state-of-the-art baseline algorithms, which operate on conventional IV-OCT images. The additional polarization contrast available to PS-OCT affords improved segmentation across a wide range of atherosclerotic lesion types and significantly improves the segmentation of the media boundaries in diseased vessels. Intravascular polarimetry with automated segmentation could be used for refined lesion characterization and may simplify and improve guidance of percutaneous coronary interventions.

\section*{Conflicts of Interest}
Drs. Villiger and Bouma are inventors on patents for OCT technology and methods that are owned by Massachusetts General Hospital and licensed to Terumo Corporation. 
Dr. Libby is an unpaid consultant to, or involved in clinical trials for Amgen, AstraZeneca, Baim Institute, Beren Therapeutics, Esperion Therapeutics, Genentech, Kancera, Kowa Pharmaceuticals, Medimmune, Merck, Norvo Nordisk, Novartis, Pfizer, and Sanofi-Regeneron. Dr. Libby is a member of the scientific advisory board for Amgen, Caristo Diagnostics, Cartesian Therapeutics, CSL Behring, DalCor Pharmaceuticals, Dewpoint Therapeutics, Eulicid Bioimaging, Kancera, Kowa Pharmaceuticals, Olatec Therapeutics, Medimmune, Novartis, PlaqueTec, TenSixteen Bio, Soley Thereapeutics, and XBiotech, Inc. 
Dr. Libby’s laboratory has received research funding in the last 2 years from Novartis. Dr. Libby is on the Board of Directors of XBiotech, Inc. Dr. Libby has a financial interest in Xbiotech, a company developing therapeutic human antibodies. Dr. Libby's interests were reviewed and are managed by Brigham and Women's Hospital and Partners HealthCare in accordance with their conflict-of-interest policies. 
Dr. Peter Libby has a financial interest in TenSixteen Bio, a company targeting somatic mosaicism and clonal hematopoiesis of indeterminate potential (CHIP) to discover and develop novel therapeutics to treat age-related diseases. 
Dr. Libby’s interests were reviewed and are managed by Brigham and Women’s Hospital and Mass General Brigham in accordance with their conflict of interest policies. 
Dr. Libby has a financial interest in Soley Therapeutics, a biotechnology company that is combining artificial intelligence with molecular and cellular response detection for discovering and developing new drugs, currently focusing on cancer therapeutics. 
Dr. Libby’s interests were reviewed and are managed by Brigham and Women’s Hospital and Mass General Brigham in accordance with their conflict of interest policies. 
Dr. Libby receives funding support from the National Heart, Lung, and Blood Institute (1R01HL134892, 1R01HL163099-01, and 1R01HL163099-01), the American Heart Association (18CSA34080399), the RRM Charitable Fund, and the Simard Fund. 

\bibliographystyle{IEEEtran}
\bibliography{ref} \label{sec:bibliography}

\end{document}